\documentclass[12pt,preprint]{aastex}
\shorttitle{Quasi-incompressible planet}
\shortauthors{Seidov }
\begin{document}
\title{The quasi-incompressible planet: some analytics}
\author{Zakir F. Seidov}
\affil{Research Institute, College of Judea and Samaria, Ariel 44837, Israel}
\email{zakirs@yosh.ac.il}
\date{\today}
\begin{abstract}
Exact and approximate analytical formulas are derived for the
internal structure and global parameters of
the spherical non-rotating quasi-incompressible planet.
The planet is modeled by a polytrope with
a small polytropic index $n \ll 1$, and solutions of the relevant differential
equations are obtained analytically,
to the second order of $n$.
Some solved and unsolved problems of polytropic models are discussed as well.
\end{abstract}
\section{Introduction}
Many classical problems of the theoretical astrophysics: internal
structure, effects of rotation, tidal effects,  pulsations
and stability of the stars/planets
are solved mainly in the case of incompressible (or homogeneous) liquid.\\
The point is that even the simplest (and basic) differential equation of internal
structure - Lane-Emden equation (hereafter LEE) of index $n$
is solved analytically only for
three values of $n=0, 1, 5$, each of them having their own deficiencies - case
$n=0$ corresponds to {\em incompressible} liquid, polytropic model with $n=1$ has
{\em constant} radius independent of its mass, and the star with $n=5$ (and with
$n>5$) has {\em infinite} radius. Some 25 years ago SK
( Seidov \& Kuzakhmedov, 1977 (SK77), 1978 (SK78))
had presented the new analytical solutions of the LEE for
 index $n$ only slightly differing from 0, 1,
and 5, and also the series
form of solution  of LEE of arbitrary index.
Several authors
 used and extended the approach
of SK for the non-rotating and slowly rotating models -
\citet{Seid78a,Seid78b,Seid79a,Seid79b,SSK79,
MoAb80,Jabb84,Caim87,Hord87,Hord90,MoRy01}.\\
In this note I return to SK and present
some analytical results for the {\em quasi-incompressible} model of planet.
The  idea is to use the {\em perturbation theory}  of differential equations:
if we have the analytical solution of the ordinary differential equation
(ODE) for some particular value of  parameter $n=n_{0}$ (say $n_{0}=0$) then we
may try to look for the analytical solution of the same ODE with
$n\approx n_{0}$ (in our case $n\approx 0$).
\section{Basic equation}
The  basic equation is LEE of index $n$ :
\begin{equation}
\frac{1}{x^{2}}\frac{d}{d\, x}\left( x^{2}\,\frac{d\, y}{d\, x}\right) =-y^{n},
\label{LEE} \end{equation}
with initial conditions $y(0)=1,\quad y^{'}(0)=0$.\\
We look for solution $y(x)$ in an interval
from $x=0$ to the first zero $x=X$ such that $y(X)=0$
(hereafter to be concise I use shorthand $s\equiv \sqrt{6}$).\\
Three classical analytical solutions of eq. (\ref{LEE}) are long
known (see e.g. \citep{Chan57}):
\begin{equation}
n=0,\quad y=1-\frac{1}{6}\,x^{2},\quad X=s,\quad \mu =2\,s,\quad
\rho_c/\rho_m=1;
\label{LEE0}\end{equation}
\begin{equation}
n=1,\quad y=\frac{\sin x}{x},\quad X=\pi,\quad \mu =\pi,\quad
\rho_c/\rho_m=\pi^{2}/3;
\label{LEE1}\end{equation}
\begin{equation}
n=5,\quad y=(1+\frac{1}{3}\,x^{2})^{-1/2},
\quad X\rightarrow \infty ,\quad \mu =\sqrt{3},\quad
\rho_c/\rho_m\rightarrow\infty.
\label{LEE5}\end{equation}
In these equations, $\mu=-X^{2}y^{'}(X),\quad \rho_c/\rho_m=X^{3}/3\,\mu$; $X$, $\mu$
are dimensionless radius and mass, and $\rho_c/\rho_m$ is the central-to-mean density
ratio.
\section{The perturbation method}\label{pert}
Consider eq. (\ref{LEE}) as ODE depending on parameter $n$,
then assuming $n$ as a small parameter, $n\ll 1$, we expand the r.s. of
eq. (\ref{LEE})
to the second order of $n$:
\begin{equation}
y = y_{0} + n\,  y_{1}+ n^{2}\,  y_{2},\quad
y^{n} = 1 + n\, \ln\, (y_{0}) + n^{2}\, \left( {y_{1}\over y_{0}}+  {1\over 2}\,\ln^{2}
(y_{0})\right).
\label{rsn} \end{equation}
From  eqs. (\ref{LEE}, \ref{rsn}) we have three coupled ODEs for three functions $y_{0},\,
y_{1},\, y_{2}$:
\begin{equation}
\frac{1}{x^{2}}\frac{d}{d\, x}\left( x^{2}\,\frac{d\, y_{0}}{d\, x}\right) =-1,
\quad y_{0}(0)=1,\quad y_{0}^{'}(0)=0;
\label{eq0} \end{equation}
\begin{equation}\frac{1}{x^{2}}\frac{d}{d\, x}\left( x^{2}\,\frac{d\, y_{1}}{d\, x}\right) =
- \ln\,(y_{0}), \quad y_{1}(0)= y_{1}^{'}(0)=0;
\label{eq1} \end{equation}
\begin{equation}\frac{1}{x^{2}}\frac{d}{d\, x}\left( x^{2}\,
\frac{d\, y_{2}}{d\, x}\right) =
-\left({y_{1}\over y_{0}}+  {1\over 2}\,\ln^{2}\,(y_{0})\right),
\quad y_{2}(0)= y_{2}^{'}(0)=0.
\label{eq2} \end{equation}
Initial conditions in eqs. (\ref{eq0}, \ref{eq1}, \ref{eq2})
are defined by the form of series expansion of the solution
of LEE of arbitrary $n$ at $x=0$:
\begin{equation}
y=1-\frac{1}{6}\,x^{2}+\frac{n}{5!}\,x^{4}+\frac{n\,(8\,n-5)}{3\cdot7!}\,x^{6}+
\frac{n\,(122\,n^2-183\,n+70)}{9\cdot9!}\,x^{8}+\ldots.
\label{ys} \end{equation}
Expanding eq. (\ref{ys}) to the second order of $n$, we have the series
expansions for functions $y_{1},\,y_{2}$ at $x=0$:
\begin{equation}\label{y1ser}
  y_{1}=\frac{1}{5!}\,x^{4}+{5\over 3\cdot7!}\,x^{6}+{70\over
9\cdot9!}\,x^{8}+\ldots ,
\end{equation}
\begin{equation}\label{y2ser}
y_{2}=-\frac{8}{3\cdot 7!}\,x^{6}-{183\over 9\cdot9!}\,x^{8}-\ldots .
\end{equation}
Note that $  y_{1} > 0$ and  $  y_{2} < 0$.\\
Before solving eqs.  (\ref{eq1}, \ref{eq2}) I'd like to dwell some on
the validity of the approach used.
There are some more or less rigorous theorems in mathematical analysis about
dependence of ODE's solutions on parameter
(see e.g. \cite{KoKo68}, sect. 10.2-7c, and \cite{Kamk59}, sect. 2.5., SK78
and references therein)
and this note is not the right place to discuss
them, see e.g. SK78. In our case, we loosely reduce these
theorems to assertion that, if for $n=0$ and $n\ll 1$ the
solutions of ODE exist, then the formal expansion of ODE and its solution in
the above mentioned way converge to the correct solution
(actually, this is rewording of the  Poincar$\acute{e}$-Lyapunov theorem,
see SK78 and references therein). In our case
 of LEE we  "do know" that desired solution exists (and we may find it e.g. by
the numerical integration of ODE) therefore we may conclude that the procedure should give
(in principle - for infinite number of terms in eq. (\ref{rsn})!) correct solution.
Does the particular truncated approximation give correct
 (at least in the numerical sense) solution  -
this is not guaranteed  and should be checked each time separately.\\
As it happens in our case, one meets no trouble with the perturbation method
in the first approximation, to first order of $n$ (see however sect. \ref{distr}),  while
already in the second approximation there are some difficulties.
\section{First approximation}\label{sol1}
First approximation (zero'th approximation coincides with LEE of index $n=0$
and needs no additional solving)
was solved in SK78 and we present here results with some
additional formulas needed for the second approximation.\\
We have from eq. (\ref{eq0}) (I remind that the shorthand $s\equiv \sqrt{6}$ is used):
\begin{equation}
y_{0}(x)=1-\frac{1}{6}\,x^{2};
\,\,y_{0}({s})=0;
\,\,y_{0}^{'}({s})=-\frac{1}{3}\,{s};
\,\,y_{0}^{''}({s})=-\frac{1}{3}.
\label{y0} \end{equation}
Using eqs. (\ref{eq0}, \ref{y0}) we get from eq. (\ref{eq1}):
\begin{equation}
y_{1}^{'}(x)=-
{1\over x^{2}}\,\int_{0}^{x}\,t^{2}\,\ln (y_{0}(t))\,dt,
\label{y1pa} \end{equation}
or
\begin{equation}\label{y1pb}
  y_{1}^{'}(x)=\frac{4}{x}+\frac{2}{9}\,x-
        \frac{1}{x}\,\ln\, (y_{0})-
       \frac{2\,s}{x^{2}}\,\mbox{arctanh}\,(\frac{x}{s}).
\end{equation}
We need also value of $y_{1}^{'}$ at $x=s$:
\begin{equation}
y_{1}^{'}(s)=\frac{2}{9}\,s\,(4-3 \ln2).
\label{y1psq6} \end{equation}
Using eq. (\ref{y1pb}) we get from eq. (\ref{eq2}):
\begin{equation}
y_{1}(x)=\int_{0}^{x}\,y_{1}^{'}(t))\,dt,
\label{y1a} \end{equation}
or
\begin{equation}
y_{1}(x)=  \frac{5\,x^2}{18}-4 +\left( 2+y_{0} \right) \,
   \ln({y_0})
   + 4\,\frac {s}{x}\,\mbox{arctanh}\,(\frac{x}{s}),
     \label{y1b} \end{equation}
\begin{equation}
y_{1}(s)=  4 \ln2 -{7\over 3}.
     \label{y1sq6} \end{equation}
\section{Second approximation}\label{sol2}
We get from eq. (\ref{eq2}):
\begin{equation}
y_{2}^{'}(x)=-
{1\over x^{2}}\,\int_{0}^{x}\,t^{2}\,
\left[{y_{1}(t)\over y_{0}(t)}+  {1\over 2}\,\ln^{2} \left(y_{0}(t)\right)\right]\,dt.
\label{y2pa} \end{equation}
Using $y_{0}(t)$ from eq. (\ref{y0}) and $y_{1}(t)$ from eq. (\ref{y1b}) we can
integrate eq. (\ref{y2pa}) analytically however expression is cumbersome and we
do not present it here. We only mention that $y_{2}^{'}(x)$ has a log-singularity at
$x=s$:
\begin{equation}
\lim_{x\rightarrow s}y_{2}^{'}(x)=-
{1\over 6}\,\int\,6\,
{y_{1}(s)\over 2\,(1-{x/s})}
\,dx={s\over 2}\,y_{1}(s)\,\ln (1-{x\over s}).
\label{y2psq} \end{equation}
Using solution $y_{2}^{'}(x)$ we get further:
\begin{equation}
y_{2}(x)=\int_{0}^{x}\,y_{2}^{'}(t)\,dt.
\label{y2a} \end{equation}
Again analytical solution is very cumbersome and we present
 only the value of $y_{2}(x)$
at $x=s$:
\begin{equation}
y_{2}(s)=\frac{1}{9}\,\left(413-21\,\pi^{2}-402\,\ln(2)+144\,\ln^{2}(2)\right).
\label{y2sq6} \end{equation}
\section{Series solution} \label{series}
In SK77 the series solution was given for LEE of arbitrary $n$:
\begin{equation}
y(x)=1+\sum^{\infty}_{k=1}\,a_{k}\,x^{2\,k},
\label{yser} \end{equation}
with recursion relation between coefficients $a_{k}$:
\begin{equation}
a_{m+1}={1\over m(m+1)(2\,m+3)}\,\sum^{m}_{i=1}\,
(i\,n+i-m)(m-i+1)\left(3+2\,(m-i)\right)\,
a_{i}\,a_{m+1-i}, \label{rec} \end{equation}
where $m\geq 1$, and $a_{1}=-1/6$.
 This recursion relation is valid for any $n$. The important problem
of the convergence radius $r$ of the series (\ref{yser}, \ref{rec})
can not be solved in the general case.\\
We consider here the interesting case of small values of $n$. In the linear
approximation, we have for the first several coefficients:
\begin{equation}
a_{1}=-{1\over 3!},\quad a_{2}={n\over 5!},\quad a_{3}={5\,n\over 3\cdot7!}
,\quad a_{4}={70\,n\over 9\cdot 9!},\quad a_{5}={3150\,n\over 45\cdot 11!}.
\label{smalln} \end{equation}
We note that starting with $a_{2}$ all coefficients are positive. To derive
a general formula for coefficients (in the same approximation,
 to the first order of $n$), we retain in the sum in r.s. of the recursion
relation (\ref{rec}) only first (with $i=1$) and last (with $i=m$) terms,
and we get the following relation:
\begin{equation}\label{am1am}
 a_{m+1}={(m-1)\,(2\,m+1)\over 6\,(m+1)\,(2\,m+3)}\,a_{m},\quad m\geq
 2,\quad a_{1}=-{1\over 6},\quad a_{2}={n\over 5!}.
\end{equation}
From this we get for the convergence radius of the series solution
of the LEE for quasi-incompressible planet:
\begin{equation}\label{rconv}
r=\lim_{m\rightarrow \infty} \sqrt{{a_{m}\over a_{m+1}}}=s.
\end{equation}
\section{Density and pressure distribution}\label{distr}
Besides solving LEE (which is of its own interest and import) it's interesting
to look for distribution of some physical parameters, namely density and
pressure over the model's volume. We mention that under the gravitationally equilibrium
 condition, the pressure distribution is more strictly defined: the pressure should be
a {\em continuous} function of radius (while density can have {\em discontinuity}  -
as in the first order phase transition) and  the pressure should
 be monotonically decreasing.\\
 Consider the spherical polytrope of index $n$,
 with central pressure $P_{c}$, and central density $\rho_{c}$.\\
 Transition from the "physical" current radius (distance from the center),
 $r$, to the "dimensional" radius $x$ (which appears in all equations above) is as
 follows:
\begin{equation}\label{trans}
 r=\alpha\cdot x,\quad
   \alpha=\left[{(n+1)\,P_{c}\over 4\,\pi\,G\,\rho_{c}^{2}}\right]^{1/2},
   \end{equation}
where $G$ is the gravitational constant.\\
Pressure $\,P(r)$ and density  $\rho\,(r)$ at distance $r$ are given by:
\begin{equation}\label{Pror}
   P\,(r)/P_{c}=\,y^{1+n}\,(x),\quad \rho\,(r)/\rho_{c}=y^{n}\,(x) ,
\end{equation}
 where $y\,(x)$ is the function which appears in all  equations above.\\
 In the incompressible planet
 (polytropic index $n=0$), the density $\rho$ is constant
 over the volume, while the pressure decreases as:
 \begin{equation}\label{Pres}
  P(r)=P_{c}\cdot y_{0}\,(x)=P_{c}\cdot(1-(x/s)^{2}).
\end{equation}
 In the quasi-incompressible planet, in the first approximation (linear in
 $n$) for density as a function of radius we have:
 \begin{equation}\label{dens1}
  \rho\,(r)/\rho_{c}=y^{n}(x)=1+n\,\ln\, (y_{0}(x))=1+n\,\ln\,(1-(x/s)^{2}),
\end{equation}
 and at $x=s$ (inside the planet close to its surface) there is a
 log-singularity  of density. This is the reducible singularity, that
 is by integrating over volume we
 get correct values of radius, mass, and central-to-mean density ratio. Still
 this singularity in the density distribution is the unpleasant and non-physical
 characteristic of the perturbation method. Sure, in reality there is
  no density singularity inside the quasi-incompressible planet, and we may avoid
  it in our calculation keeping $n$ in power of $y(x)$
  and rewrite eq. (\ref{dens1}) as follows:
   \begin{equation}\label{dens2}
\rho\,(r)/\rho_{c}=y^{n}(x)=[y_{0}(x)+n\,y_{1}(x)]^{n},
\end{equation}
or even more accurately:
   \begin{equation}\label{dens3}
\rho\,(r)/\rho_{c}=y^{n}(x)=[y_{0}(x)+n\,y_{1}(x)+n^{2}\,y_{2}(x)]^{n}.
\end{equation}
 As to the pressure distribution, there
 is  no difficulty because in the linear approximation we have:
 \begin{equation}\label{Pres1}
 P(r)/P_{c}=y^{n+1}(x)=
  y_{0}(x)+n\,\left[y_{1}(x)+y_{0}(x)\,\ln\,(y_{0}(x))\right],
\end{equation}
and there is no singularity, in the pressure distribution, inside the model
 (while there is a log-singularity in {\em derivative} $dP/dr$ which is proportional
 to the density - in accordance with eq. (\ref{dens1})).\\
Again, as for the density distribution, we may write more accurately:
 \begin{equation}\label{Pres2}
 P(r)/P_{c}=[y_{0}(x)+n\,y_{1}(x)+n^{2}\,y_{2}(x)]^{n+1},
\end{equation}
avoiding the singularity in the derivative $dP/dr$.
\section{Global parameters} \label{glob1}
Now we are ready to find some global parameters of the quasi-incompressible
planet.
\subsection{Radius} \label{X1}
According to eq. (\ref{trans}), the total radius $R$ of the polytrope is
 \begin{equation}\label{torad}
R = \alpha \cdot X,
\end{equation}
where the first zero $X$ is such that:
\begin{equation}\label{eqX}
 y\,(X)= y_{0}(X)+n\,y_{1}(X)+n^{2}\,y_{2}(X)=0,
\end{equation}
\begin{equation}\label{Xn2}
 X=s+n\, \delta_{1}+n^{2}\, \delta_{2},
\end{equation}
with $\delta_{1}$, $\delta_{2}$ still to be found.\\
In the spirit of the perturbation method, we should keep, in eq. (\ref{eqX}),
the argument $X$ in $ y_{0}(X)$ to the second order of $n$,
in $ y_{1}(X)$, to the first order of $n$,
and in $ y_{2}(X)$,  only $X=s$.\\
Expanding all functions to the second order of $n$, combining terms with the same power
of $n$, equalizing  coefficients of $n$ and $n^{2}$ to zero, we get system of
two equations for $\delta_{1}$ and $\delta_{2}$, and solving these
equations we get finally:
\begin{equation}\label{delt1}
 \delta_{1}={3\over s}\,y_{1}(s)={1\over s}\,(12\,\ln\,2 -7);
\end{equation}
\begin{equation}\label{delt2}
 \delta_{2}={3\over s}\left(y_{2}(s)+ {3\over s}\,
 y_{1}(s)\,y_{1}^{'}(s)-{1\over 4}\,y_{1}^{2}(s)\right),
\end{equation}
\begin{equation}\label{delt2a}
 \delta_{2}=\frac{1}{12\,s}\,
 \left(1379 - 84\,\pi^{2}- 888\,\ln\,2 + 144\,\ln^{2}\,2\right).
\end{equation}
From eqs. (\ref{Xn2}, \ref{delt1}, \ref{delt2a}) we have numerically:
\begin{equation}\label{X12num}
 X=2.44948974278+0.537975784794\,n+0.12328309\, n^{2}.
\end{equation}
Note that $X$ (and hence radius $R$) is the only parameter
of the quasi-incompressible planet which
can be calculated to the second order of $n$. Due to the singularity of
$y_{2}^{'}(s)$, the second approximation is not possible for
other global parameters of the quasi-incompressible planet.
\subsubsection{Comparison with the numerical values} \label{com1}
For $n=1/2$, numerically $X_{num}=2.752698$, analytically, from eq. (\ref{X12num})
 $X=2.749298$ with relative difference of about -.12\%.\\
For $n=1/10$, numerically $X_{num}=2.504545$, analytically $X=2.50452015$ with relative
difference of .01\%.\\
For $n=1/1000$, numerically $X_{num}=2.45004$, analytically $X=2.45002978$ with
difference in the last digit of $X_{num}$.\\
We mention that the cases of very small
values of $n$ are more easier (and more accurately) described by analytical
formulas than by numerical calculations.\\
Besides, no numerical calculation can give
analytical formula for $X(n)$ even in the linear approximation in $n$.
\subsection{Mass} \label{mu}
The running mass inside the polytropic sphere of radius $r$ is:
 \begin{equation}\label{runma}
 m(r) = 4\,\pi\,\rho_{c}\, \alpha^{3}\,[-x^{2}\,y^{'}(x)],
\end{equation}
and the total mass of the polytrope is:
 \begin{equation}\label{toma}
 M = 4\,\pi\,\rho_{c}\, \alpha^{3}\,\mu,\quad \mu=-X^{2}\,y^{'}(X).
\end{equation}
In order to calculate  $\mu(X)$ to the second order of $n$,
 we need $y_{2}^{'}(s)$ which is infinite (see eq. (\ref{y2psq})),
 so we restrict ourselves by the linear approximation,  already
considered in SK78. By the same way as in section (\ref{X1}), we get:
\begin{equation}\label{mun1}
  \mu=2\,s+6\,n\,\left[\delta_{1}-y_{1}^{'}\,(s)\right]=2\,
  s\,\left[1-(37/6-8\,\ln\,2)\,n\right]=2\,s\,(1-0.62148922\,n).
\end{equation}
\subsection{Central-to-mean density ratio} \label{rorom}
For the polytrope, from eqs. (\ref{torad}, \ref{toma}) we have for
the central-to-mean density ratio:
\begin{equation}\label{rorom1}
{\rho_c\over \rho_m}={X^{3}\over 3\,\mu},
\end{equation}
and though we calculated $X$ in the second approximation, however the parameter  $\mu$
was calculated only in the first approximation,
therefore we can calculate the central-to-mean density ratio
for the quasi-incompressible planet
only in the first approximation, already considered in SK78:
\begin{equation}\label{rorom2}
{\rho_c\over \rho_m}=1+\left({8\over 3}-2\,\ln \,2\right)\,n=
  1+1.280372\,n.
\end{equation}
\subsection{Moment of inertia}\label{inert}
The central moment of inertia of the spherical polytrope is:
\begin{equation}\label{momin}
I=\int_{0}^{M}r^{2}\,d\,m=k_{I}\,M\,R^{2},\quad  k_{I}={1\over \mu\,X^{2}}\,\int_{0}^{X}\,y^{n}(x)\,x^{4}\,d\,x,
\end{equation}
and to the first order of $n$ we get:
\begin{equation}\label{momina}
k_{I}= \frac{3}{5}-{6\,n\over 25}.
\end{equation}
\subsection{Milne integral}\label{Milne}
In the theory of polytropes, there is Milne's relation \citep{Miln29} which
also appeared in the "refined" theory of rotating polytropes
(\citep{ChLe62}, eq. (87)):
\begin{equation}\label{miln}
 MI= \int_{0}^{X}\,y^{1+n}(x)\,x^{2}\,d\,x={1+n\over 5-n}\,X^{3}\,\left[y^{'}(X)
\right]^{2}.\end{equation}
Using our formulas we find for both sides of the equation (\ref{miln}), for the
quasi-incompressible planet ($n<<1$):
\begin{equation}\label{miln1}
MI={4\over 5}\,s+{2\over 25}\,s\,(120\,\ln\,2-79)\,n.
\end{equation}
We can not calculate $MI$ in the second approximation because this requires
the $y_{2}^{'}(s)$ (see r.s. of  eq. (\ref{miln})) which diverges (see
eq. (\ref{y2psq})).\\
By the way, the case of $n=5$ in eq. (\ref{miln}) is interesting as well.
Assuming $(5-n0$ as a small parameter, expanding
both sides of eq. (\ref{miln}) in series we find the dependence of $X(n)$,
already found in SK78:
\begin{equation}\label{Xn5}
X(n)={32\,\sqrt{3}\over \pi}\,{1\over 5-n},\quad 0 < 5-n \ll 1.
\end{equation}
\subsection{One interesting relation} \label{intrel}
In \citep{SSkv00}, the interesting relation is introduced
between  the gravitational potential energy, $W$, the central potential, $U_{c}$,
and the mass of the celestial body.
We confine ourselves  here by the case of spherical polytropes.\\
From the theory of polytropes, the central gravitational potential, $U_{c}$,
the central pressure $P_{c}$, and the central density, $\rho_{c}$,
are (see \citep{Chan57}, p. 100, eq. (85); p. 99, eq. (80,81);  p. 78, eq. (99)):
\begin{equation}\label{Uc1}
 U_{c}=(1+n)\,{P_{c}\over \rho_{c}}+{G\,M\over R},
\end{equation}
\begin{equation}\label{Pc}
P_{c}={1\over 4\,\pi\,(1+n)\,[y^{'}(X)]^{2}}\,{G\,M^{2}\over R^{4}},
\end{equation}
\begin{equation}\label{roc}
\rho_{c}={1\over 3}\,{X\over [-y^{'}(X)]}\,\rho_{m}={X\over [-y^{'}(X)]}
{M\over 4\,\pi\,R^{3}}.
\end{equation}
Combining eqs. (\ref{Uc1}, \ref{Pc}, \ref{roc}), we get for
the central gravitational potential of the spherical polytrope:
\begin{equation}\label{Uc2}
 U_{c}=  \left(1+{1\over [-X\,y^{'}(X)]}\right)\,{G\,M\over R}.
\end{equation}
Potential energy of polytrope (we loosely take positive sign) is
(\citep{Chan57}, p. 101, eq. (90)):
\begin{equation}\label{W}
 W= {3\over 5-n}\,{G\,M^{2}\over R}.
\end{equation}
Now we have, for polytropes, the {\em WUM}-ratio introduced in \citep{SSkv00}:
\begin{equation}\label{WUM}
 WUM ={W\over U_{c}\,M}= {3\over 5-n}\,{1\over 1+X/\mu}.
\end{equation}
Note in brackets that $WUM$-ratio is remarkably constant (=2/5)
for the rotating triaxial
homogeneous ellipsoids (see \citep{SSkv00}).
In our case of the quasi-incompressible planet,
using eqs. (\ref{Xn2}, \ref{delt1}, \ref{mun1}, \ref{WUM}) we have to the first order of $n$:
\begin{equation}\label{WUMn1}
 WUM = {2\over 5}\,\left[1-n\,({22\over 15}- 2 \,\ln 2)\right].
\end{equation}
\subsection{Parameter $\omega$} \label{omega}
In \citep{CrMu94}, the parameter $\omega$ for the polytrope
was introduced (see their eq. (A14)):
\begin{equation}\label{omegacdm}
  \omega={1\over 4 \pi (1+n)}\left({3\over X}\right)^{1+1/n}\left[-y^{'}
(X)\right]^{1-1/n},
\end{equation}
 with a note that at $n=0$ this expression is undefined, and from numerical
 calculations, at $n=0$, $\omega\approx 0.033175949$.
 This is a good chance to demonstrate usefulness of the perturbation method by SK:
 using the formulas of the previous sections and omitting some algebra
  we write down the final expression for $\omega$  to the first order of $n$:
\begin{equation}\label{omegaSK}
  \omega={3\over 2 \pi}\,\exp(-8/3)\,
\left\{1-n\left[1+{1\over 2}\left({8\over 3}-\ln 6\right)^{2}\right]\right\},
\end{equation}
or numerically
$$ \omega=0.033175904175\, (1. - 1.382731302\, n),$$
and we note again that the analytical expression helps to check the numerical
calculations (and the last two digits of the "numerical" value of $\omega$ at $n=0$
 differ from the "analytical" value).
\section{"Numerical" perturbation method}
Though being slightly out-of-topic, we mention that the perturbation method
can be used
in its numerical modification as well. \\
In the case of
polytropes, it is particularly interesting to do this for the case $n=3$.
We briefly describe the method and give the short results in the first
approximation. We take in the r.s. of eq. (\ref{LEE}), $n=3-\delta$ with
$0 < \delta \ll 3$, then we have
\begin{equation}\label{rsn3d}
\left(y_{0}-\delta \,y_{1}\right)^{3-\delta}=
y_{0}^{3}-y_{0}^{2}\,(3\,y_{1}+y_{0}\,\ln\,y_{0})\,\delta.
\end{equation}
Here $y_{0}$ is solution of LEE of index $n=3$ and $y_{1}$ is the "perturbation"
function. Also, because $ \delta > 0$, zero $X$ of function $y_{0}-\delta \,y_{1}$
is {\em less} than $X_{0}$, zero of function $y_{0}$.
Due to an inequality $X < X_{0}$
 there is no problems with behavior of functions near boundary as
in the quasi-incompressible case. As in the text above, we have system of two ODE's
(each of them of the second order) for functions $y_{0}$ and $y_{1}$.
We can solve this system  numerically and find all relevant functions $y_{0}$, $y_{1}$,
$y_{0}^{'}$, and $y_{1}^{'}$ and particularly their values at $x=X_{0}$. Then
we can get the formulas similar to ones in section (\ref{glob1}). \\
We present the results of the numerical calculations:
\begin{equation}\label{X0y0p3}
  X_{0}=6.8968498,\quad y_{0}^{'}(X_{0})=-0.0424297317,
\end{equation}
\begin{equation}\label{y13y1p3}
  y_{1}(X_{0})=0.16547670449,\quad y_{1}^{'}(X_{0})=-0.00616735.
\end{equation}
From this we have for the dimensionless radius of polytrope
with index $n$ close to 3:
\begin{equation}\label{Xn31st}
 X=6.896848+(3-n)\,y_{1}(X_{0})/y_{0}^{'}(X_{0})=6.896848-3.90002\,(3-n).
\end{equation}
Numerically, at $n=3-0.1$, $X=6.526374$, while from the linear perturbation method
(\ref{Xn31st}): $X=6.5068$.\\
Also, numerically, at $n=3+0.1$, $X=7.308484$,
 while from the linear perturbation method (\ref{Xn31st}): $X=7.2868$.\\
 Deviations of numerical values  in two cases
 correspond (correctly) to the positive second derivative of function $X(n)$.
\section{Summary}
In this  note we present the exact analytical solutions for the internal structure
and global parameters of the quasi-incompressible planet modeled as the
polytrope of small index $0 < n \ll 1$. The  perturbation method used here is
not rigorously justified by means of the theory of differential equations and there
are some problems about application of the method in the interval of argument
 where the perturbation function is of the same order or
 even larger than the initial non-perturbed function.
 The problem of justification is here similar to the problem of rotationally
 distorted polytropes which was already discussed some
 in the astrophysical literature (see e.g. \cite{ChLe62}).
 Still validity of any method in applicational sciences (as astrophysics is such
 relative to mathematics) may be compared by numerical calculation
 and (astro)-physical "common sense" and
 I only hope that this note may trigger some discussion as well.
\section{Acknowledgments}
 This paper is to be devoted to the memory of my dear friend and
 PhD student late Rafael Khejrullaevich Kuzakhmedov (=Rafik) who having been a very
 strong theor. physicist flatly rejected to admit any significance in our two minor
 notes (SK77, SK78) and had never got even PhD degree. Only my vision of
 astrophysics and Rafik's incredible ability of what is called now symbolic
mathematics (and my long pressure on Rafik!) could eventually produce
these two short notes which I count among my best ones.\\
Also the financial support from the Israeli Ministry of Science
 and the Administration of the College of Judea and Samaria is duly acknowledged.


\newpage
\begin{thebibliography}{}
\bibitem[Caimmi(1987)]{Caim87} Caimmi, R.  1987, \apss, 140, 1
\bibitem[Chandrasekhar(1957)]{Chan57} Chandrasekhar, S. 1957,
 An Introduction to the Study of Stellar Structure (Chicago: Dover)
\bibitem[Chandrasekhar \& Lebovitz (1962)]{ChLe62}
Chandrasekhar, S., \& Lebovitz, N. R. 1962, \apj, 136, 1082
  \bibitem[Christensen-Dalsgaard \& Mullan(1994)]{CrMu94}
 Christensen-Dalsgaard, J., \& Mullan, D. J. 1994, \mnras,  270, 921
\bibitem[Horedt(1987)]{Hord87} Horedt, G. P.  1987,
Astron. Ap., 172, 359
\bibitem[Horedt(1990)]{Hord90} Horedt, G. P.  1990, \apj, 357, 560
\bibitem[Jabbar(1984)]{Jabb84} Jabbar, J. R. 1984, \apss, 100, 447
\bibitem[Kamke (1959)]{Kamk59} Kamke, E. 1959,
Differentialgleichunge (6th ed., Leipzig)
\bibitem[Korn \& Korn (1968)]{KoKo68} Korn, G. A. \& Korn, T. M. 1968,
Mathematical Handbook (2nd ed., New York: McGraw-Hill)
\bibitem[Medvedev \& Rybicki(2001)]{MoRy01} Medvedev, M. V., \&  Rybicki, G.
2001, \apj, 555, 863
  \bibitem[Milne(1929)]{Miln29}
 Milne, E. A.  1929, \mnras,  89, 739
\bibitem[Mohan \& Al-Bayaty(1980)]{MoAb80} Mohan, C. \& Al-Bayaty, A. R. 1980,
\apss, 73, 227
\bibitem[Seidov (1978a)]{Seid78a} Seidov, Z. F. 1978,
 Sov. Astron. Lett., 4(3), 144
 \bibitem[Seidov (1978b)]{Seid78b} Seidov, Z. F. 1978,
in Sources of Gravitational Radiation, Proceed. Batelle Seattle Workshop,
 ed. L.L. Smarr (New York: CUP), 480
 \bibitem[Seidov (1979a)]{Seid79a} Seidov, Z. F. 1979,
Dokl. AN AzSSR, 35, no. 1, 21
 \bibitem[Seidov (1979b)]{Seid79b} Seidov, Z. F. 1979,
in IAU Coll. no. 53, Rochecter, (pp. 478-482)
\bibitem[Seidov \& Kuzakhmedov(1977)]{SK77}
Seidov, Z. F., \& Kuzakhmedov, R. Kh. 1977,
 Sov. Astron., 21, 399 (SK77)
\bibitem[ Seidov \& Kuzakhmedov(1978)]{SK78}
Seidov, Z. F., \& Kuzakhmedov, R. Kh. 1978,
 Sov. Astron.,  22, 711 (SK78)
 \bibitem[ Seidov, Sharma \& Kuzakhmedov(1979)]{SSK79}
  Seidov, Z. F., Sharma, J. P., \& Kuzakhmedov, R. Kh. 1979,
  Dokl. AN AzSSR, 35, no. 5, 21; no. 6, 25
   \bibitem[ Seidov \& Skvirsky (2000)]{SSkv00}
  Seidov, Z. F., \& Skvirsky, P. I. 2000, preprint (astro-ph/0003064)
\end{thebibliography}
\end{document}